\documentclass[a4paper,twocolumn,11pt]{quantumarticle}
\pdfoutput=1
\usepackage{dcolumn}
\usepackage{amsmath, amssymb}
\usepackage{graphicx}
\usepackage{braket}
\usepackage{xcolor}
\usepackage{subcaption}
\usepackage{hyperref}
\usepackage[numbers,sort&compress]{natbib}

\begin{document}
\title{Bloch-sphere picture and restoration of the quantum Mpemba effect beyond the weak-coupling regime in the spin–boson model
}

\author[1]{P. Chirico}
\email{p.chirico@ssmeridionale.it}

\author[2,3]{G. Di Bello}

\author[3,4]{G. De Filippis}

\author[3,4]{C. A. Perroni}

\affil[1]{Scuola Superiore Meridionale, Largo San Marcellino 10, I-80138 Napoli, Italy}

\affil[2]{Dip. di Fisica E. Pancini -- Università di Napoli Federico II, I-80126 Napoli, Italy}

\affil[3]{INFN, Sezione di Napoli -- Complesso Universitario di Monte S. Angelo, I-80126 Napoli, Italy}

\affil[4]{SPIN-CNR and Dip. di Fisica E. Pancini -- Università di Napoli Federico II, I-80126 Napoli, Italy}

\begin{abstract}
Understanding relaxation dynamics in open quantum systems is a central problem in nonequilibrium quantum physics. Here we investigate the quantum Mpemba effect in the spin–boson model. In the weak-coupling Markovian regime we show that the occurrence of the effect strongly depends on the choice of distance measure at low temperature: while it appears in the trace distance, it can disappear in the quantum relative entropy. Going beyond the weak-coupling approximation, numerically exact simulations of the full system–bath dynamics reveal that increasing coupling enhances the effect in the trace distance and restores it in the quantum relative entropy. We uncover a simple Bloch-sphere picture of the effect: within the
excited-state hemisphere, pairs of states related by rotations and
having the same Bloch-vector modulus generically exhibit an inversion
of their relaxation ordering. This behavior is
robust across sub-Ohmic, Ohmic, and super-Ohmic bath spectra. In the
Ohmic and sub-Ohmic regimes, the crossing time becomes strongly
suppressed upon approaching the critical region, whereas in the
super-Ohmic regime the effect remains enhanced despite the absence of
a finite-coupling localization transition. These results highlight the
interplay between geometry, distance measures, system--environment
coupling, and bath spectral properties in anomalous quantum relaxation.
\end{abstract}

\maketitle
\section{Introduction}
The Mpemba effect denotes the counterintuitive phenomenon whereby a system initially prepared farther from equilibrium relaxes faster than one that is initially closer~\cite{Mpemba1969}. Originally discussed in classical systems \cite{Lu2017}, for example colloidal ones \cite{Kumar2020}, this phenomenon is referred to in the quantum setting as the quantum Mpemba effect \cite{Ares2025}.
The Quantum Mpemba effect has recently attracted significant theoretical \cite{Entanglement, PhysRevLett.133.010401, sapui2026ergotropicmpembacrossingsfinitedimensional, Yu2025, das2025rolereversalquantummpemba} and experimental \cite{PhysRevLett.133.010402, zhang,PhysRevLett.133.010403} attention in quantum simulators, as well as in both closed \cite{Calabrese_2026} and open quantum systems \cite{Nava2024, hhgj-89gj,Longhi2025, PhysRevLett.127.060401, PhysRevLett.133.140404, PhysRevLett.131.080402}.

In the quantum setting, the characterization of relaxation requires specifying a distance measure between quantum states \cite{ali2026quantum}.
Recent analytical results have established the existence of the quantum Mpemba effect for open systems whose reduced dynamics is governed by a Lindblad master equation satisfying quantum detailed balance \cite{PhysRevLett.127.060401, PhysRevLett.133.140404}.
In this framework, the effect has been rigorously demonstrated at finite temperature focusing on the quantum relative entropy as a measure of distance. Moreover, recent studies have also shown that memory effects and non-Markovian dynamics can qualitatively modify the occurrence of the quantum Mpemba effect \cite{Strachan2025,Zhang2025Mpemba,QuantumDots2025,lejeune2026accelerating, OrtegaTaberner2026}.
These results raise two natural questions: whether the occurrence of the quantum Mpemba effect depends on the specific choice of distance measure, and whether it persists at zero temperature or can even be activated beyond the weak-coupling regime. 

Here, we address these questions by studying the relaxation dynamics of the spin-boson model, a paradigmatic model of dissipation in quantum systems.
In this model, the dissipation of the two-level system is due to the coupling to a continuum of bosonic modes, namely the bath. 
We focus on the zero-temperature limit and analyze the dynamics using both the trace distance and the quantum relative entropy.
By representing the reduced dynamics on the Bloch sphere, we uncover a
simple geometric structure underlying the effect. For initial states
within the excited-state hemisphere and with the same Bloch-vector
modulus, the state initially farther from the stationary state can
overtake the initially closer one during the relaxation. In the
weak-coupling regime, this geometrical picture can be established
analytically for the trace distance in terms of the separation between
population and coherence relaxation timescales.
A similar picture on the Bloch-sphere has recently been observed in \cite{sapui2026ergotropicmpembacrossingsfinitedimensional}, where they study the Mpemba effect using as distance measure the ergotropy. We further investigate the robustness of this behavior with respect to the bath spectral properties by considering sub-Ohmic, Ohmic, and super-Ohmic environments.

\subsection*{Main results}
We find that the quantum Mpemba effect exhibits a simple geometric
structure on the Bloch sphere and strongly depends on the choice of
distance measure in the weak-coupling regime. Beyond weak coupling,
increasing the system--bath interaction enhances the effect in the
trace distance and restores it in the quantum relative entropy. This
behavior is robust across sub-Ohmic, Ohmic, and super-Ohmic
environments. In the Ohmic and sub-Ohmic regimes, the crossing time is
strongly suppressed upon approaching the critical region, whereas in
the super-Ohmic regime a similar enhancement occurs despite the absence
of a finite-coupling localization transition. Overall, our results
provide a unified picture of the quantum Mpemba effect in the
spin--boson model, clarifying the interplay between geometry, distance
measures, coupling strength, and bath spectral properties beyond the
standard Markovian weak-coupling paradigm.

\section{Model and methods} 
In the following, we consider the spin-boson model \cite{Leggett1987} fixing the Planck constant $\hbar=1$ and Boltzmann constant $k_B=1$. The total Hamiltonian reads 
\begin{equation}
H = H_S + H_B + H_{SB},
\end{equation}
where the system Hamiltonian is
\begin{equation}
H_S = -\frac{\Delta}{2}\,\sigma_x,
\label{Hamil_S}
\end{equation}
where $\sigma_x \ket{\pm} = \ket{\mp}$. In this model, the system undergoes dissipation due to the environment. 
$\Delta$ is the tunneling parameter, which quantifies the tunneling amplitude between the states $\ket{+}$ and $\ket{-}$. In order to represent the reduced density operator $\rho $ of the two-level system, we adopt the Bloch-vector representation:
\begin{equation}
\rho = \frac{1}{2}\left( \mathbb{I} + \mathbf{r}\cdot{\mathbf{\sigma}} \right).
\end{equation}
Using this representation, the state is fixed through $\mathbf{r}$.
The bath is modeled as a collection of harmonic oscillators, with each mode labeled by the index $k$. The creation and annihilation operators $b_k^\dagger$ and $b_k$ satisfy the canonical bosonic commutation relations:
\begin{equation}
[b_k, b_{k'}^\dagger] = \delta_{kk'}, \quad [b_k, b_{k'}] = [b_k^\dagger, b_{k'}^\dagger] = 0.
\end{equation}
These operators act on the Hilbert space of the bath, and the total bath Hamiltonian is given by:
\[
H_B = \sum_k \omega_k b_k^\dagger b_k,
\]
where $\omega_k$ is the frequency associated with the $k$-th mode.

The system-bath interaction is modeled as a coupling between the two-level system and the bath modes. This interaction term is written as:
\[
H_{SB} = \sigma_z \otimes \sum_k g_k \left( b_k^\dagger + b_k \right),
\]
where $\sigma_z$ is the Pauli $z$ operator acting on the two-level system and \( g_k \) are the coupling constants between the system and the $k$-th mode of the bath. This interaction gives rise to dissipative relaxation and decoherence of the two-level system. The environment is fully characterized by the spectral density
\begin{equation}
J(\omega) = \sum_k g_k^2 \delta(\omega - \omega_k),
\end{equation}
which we take in the generalized Ohmic (power-law) form with an exponential cutoff
\begin{equation}
J(\omega) = \frac{\alpha}{2}\,\omega_c^{1-s}\omega^s \,e^{-\omega/\omega_c}.
\end{equation}
Here, $\alpha$ is a dimensionless parameter controlling the system--bath coupling strength, $\omega_c$ is the high-frequency cutoff, and $s$ is the spectral exponent characterizing the low-frequency behavior of the environment. In particular, the bath is classified as sub-Ohmic for ($0<s<1$), Ohmic for ($s=1$), and super-Ohmic for ($s>1$). In the following, we consider the representative values $s=0.2$, $s=1$, and $s=1.5$, corresponding to the sub-Ohmic, Ohmic, and super-Ohmic regimes, respectively. We fix the cutoff frequency to  $\omega_c=60\Delta$.

The equilibrium properties of the spin--boson model depend qualitatively on the spectral exponent. At zero temperature, both
the sub-Ohmic and Ohmic models can exhibit a delocalized-to-localized
quantum phase transition at a critical coupling $\alpha_c(s)$ \cite{le2010quantum, de2023signatures, di2024environment}. The value and the critical properties of this transition depend on
$s$. In particular, for the Ohmic bath one has
$\alpha_c \simeq 1$ in the scaling limit considered here. By contrast,
for a super-Ohmic bath no localization transition occurs at finite
coupling, and the system remains in the delocalized phase.

For these systems, a simple analytical description of the dynamics is not available, since it is very difficult to analytically handle all the bath degrees of freedom.
On the other hand, in the weak-coupling regime, the evolution of the reduced density matrix \(\rho(t)\) can be effectively described by a Markovian quantum master equation of Lindblad type \cite{10.1093/acprof:oso/9780199213900.001.0001}
\begin{equation}
\frac{d\rho(t)}{dt} = \mathcal{L}[\rho(t)],
\label{Master equation}
\end{equation}
with $\mathcal{L}$ the Lindbladian operator, whose steady state is the thermal Gibbs state

\begin{equation}
\rho^G_{\mathrm{ss}} = \frac{e^{-\beta H_S}}{\mathrm{Tr}\!\left(e^{-\beta H_S}\right)},
\end{equation}
with inverse bath temperature \(\beta=1/T\), which is independent of system-bath coupling.
In this work, we go beyond the weak-coupling regime by numerically determining the dynamics via tensor network techniques within the Matrix Product State (MPS) framework, where we simulate the full system state, including the bath modeled as a set of $N$ harmonic oscillators. For the ground-state search, we use the Density Matrix Renormalization Group (DMRG) algorithm, and we subsequently perform the time evolution by means of matrix-product operators \cite{SCHOLLWOCK201196,PAECKEL2019167998,DiBelloEPJP,DiBelloQST}.
The DMRG states are the asymptotic stationary states, defined as:
\begin{equation}
\rho_{ss} = \mathrm{lim}_{t\to\infty} \rho(t)
\label{limit}.
\end{equation}
We emphasize that, as soon as the coupling is switched on, Eq.\,\eqref{limit} is different from the Gibbs state determined solely by $H_S$.

\section{Results}
We first discuss the picture on the Bloch sphere, then compare the two distance measures in the weak-coupling regime, and finally address the restoration of the effect beyond weak coupling. 

The relaxation towards equilibrium is quantified by a distance measure
\(D[\rho(t),\rho_{\mathrm{ss}}]\)\,\cite{Ares2025}, 
where $\rho_{ss}$ is the stationary state in Eq.\,\eqref{limit}.
For two distinct initial states, a Mpemba effect is said to occur when the ordering of their distances from the stationary state inverts at a finite time during the relaxation dynamics \cite{Ares2025}. This definition implies that the curves of the two distances must intersect at a finite time.

In this work, we have used two distance measures with respect to the stationary state: the quantum relative entropy and the trace distance.
These are respectively defined as follows:
\begin{equation}
S\big(\rho\|\rho_{\mathrm{ss}}\big)
=
\mathrm{Tr}\!\left[
\rho\big(\ln\rho-\ln\rho_{\mathrm{ss}}\big)
\right],
\end{equation}
\begin{equation}
    D(\rho,\rho_{ss}) = \frac{1}{2}\mathrm{Tr} \Big(\sqrt{(\rho - \rho_{ss})^\dagger (\rho - \rho_{ss})} \Big) .
    \label{eq:trace_distance}
\end{equation}
In particular, the quantum relative entropy provides a natural measure of distinguishability between quantum states and plays a central role in non-equilibrium quantum thermodynamics.
On the other hand, the trace distance provides a natural metric.

Previous analytical results for the spin-boson model establish the existence of the quantum Mpemba effect only in the finite-temperature ultra-weak coupling limit \cite{PhysRevLett.127.060401, PhysRevLett.133.140404}, when the Lindblad equation is used.
In particular, considering the quantum relative entropy, it has been demonstrated that one can always implement a unitary transformation that maps an arbitrary state onto another state that is orthogonal to the slowest decaying mode and therefore exhibits a faster relaxation dynamics \cite{Kochsiek2022}. 
In the case of the spin-boson model, the optimal unitary transformation is the one that maps the state onto the excited state of $H_S$. 

\begin{figure}[t]
\includegraphics[width=\linewidth]{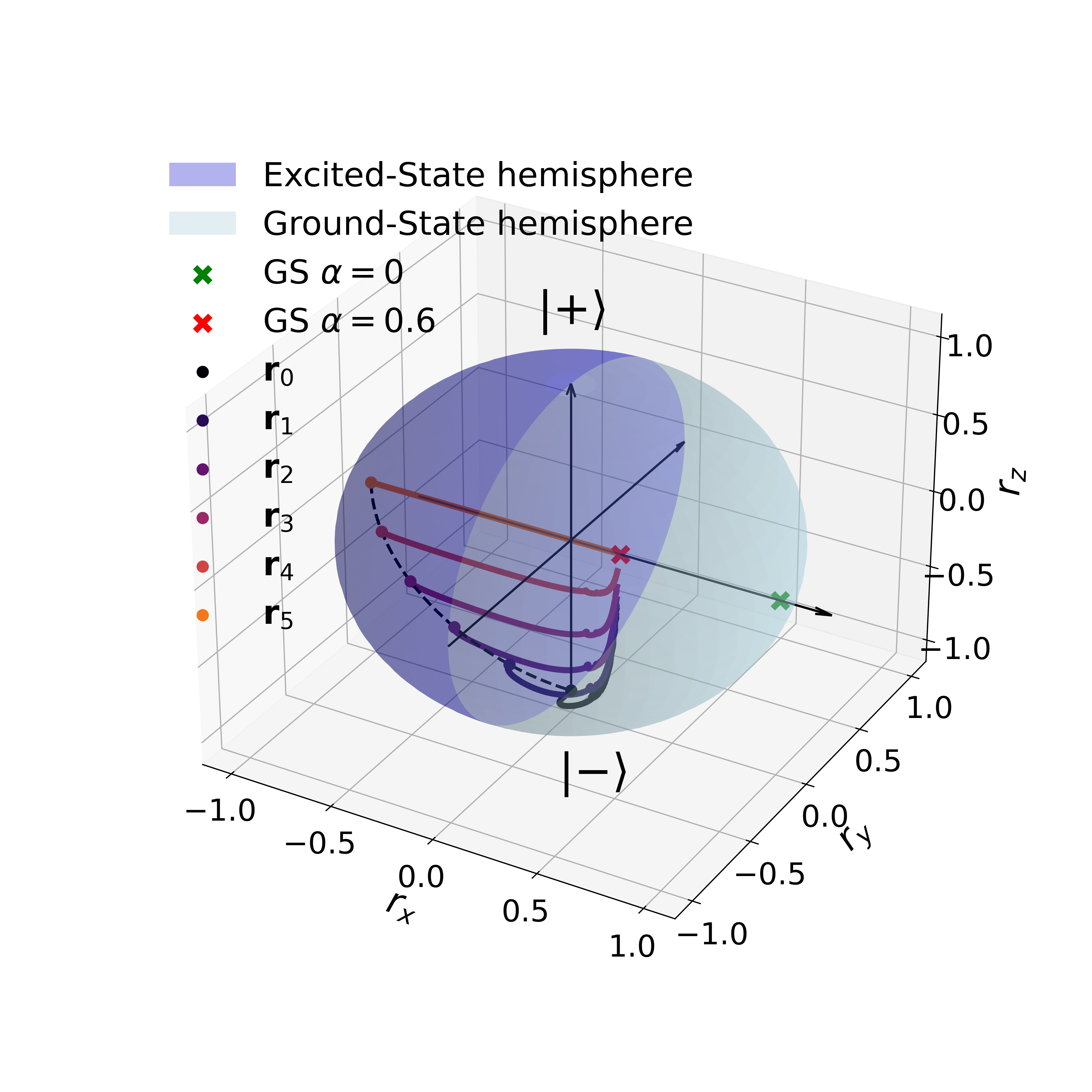}
\caption{
Geometric structure of the quantum Mpemba effect on the Bloch sphere. 
The sphere is partitioned into ground-state (GS) (light) and excited-state (dark) hemispheres. 
Colored curves represent dynamical trajectories for $\alpha=0.6$ obtained from MPS simulations for different initial states with fixed Bloch-vector modulus. 
The initial conditions used for the dynamics are $\mathbf{r_i} = (-\mathrm{sin}(\phi_i),0,-\mathrm{cos}(\phi_i))$, with $\phi_i\in \{0,0.1\pi,0.2\pi, 0.3 \pi, 0.4\pi,0.5\pi\}$.
The green and red markers denote the reduced ground states at $\alpha=0$ and $\alpha=0.6$, respectively, showing the progressive loss of purity as $\alpha \to \alpha_c^-$; in this limit, the system’s reduced ground state approaches the maximally mixed state. 
}
\label{fig:Geometric_structure}
\end{figure}

We first show analytically that the quantum Mpemba effect exhibits a simple and robust \emph{picture} when the initial states are classified according to their position relative to the energy eigenstates of the two-level system.
Using the Bloch representation of the state of the system and since $H_S=-\Delta\sigma_x/2$, the energy eigenstates lie on the x-axis of the Bloch sphere. In the following, we therefore define the ground-state and excited-state hemispheres with respect to the sign of $\langle \sigma_x \rangle$ (see Fig.\,\ref{fig:Geometric_structure}).

Focusing on initial states belonging to the excited-state hemisphere, we find that the
Mpemba effect occurs for \emph{any} pair of distinct initial states within this region related by a rotation (i.e., same $|\mathbf{r}|$).
That is, given two initial states whose Bloch vectors lie in the same excited-state hemisphere, the state initially farther from equilibrium relaxes faster and overtakes the
other at a finite time during the dynamics. 

\begin{figure}[t]
\includegraphics[width=\linewidth]{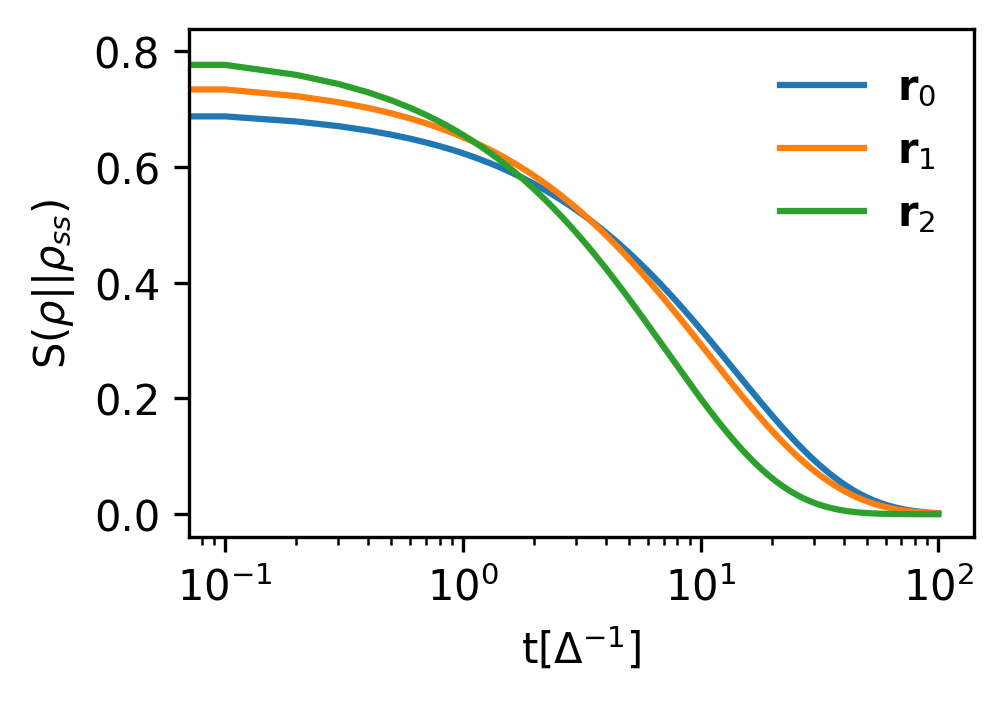}
\caption{
Quantum relative entropy with respect to the stationary state as a function of time for three initial states at high temperature ($T=5 \Delta, \Delta=1, \omega_c = 60\Delta$, $\alpha=10^{-3}$), labeled by their Bloch vectors
$\mathbf{r}_0 = (0,0,1)$, $\mathbf{r}_1=(-\frac{1}{2},0,\frac{\sqrt{3}}{2})$, and $\mathbf{r}_2=(-1,0,0)$. 
The occurrence of a quantum Mpemba effect is signaled by the finite time crossing of the curves, calculated by solving the Lindblad equation. 
}

\label{fig:bloch_mpemba}
\end{figure}

In the End Matter, we provide an analytical proof for this geometrical structure using the trace distance in the weak-coupling regime. In order to make contact with the results of Ref.\,\cite{PhysRevLett.133.140404}, we numerically demonstrate the same features by using the quantum relative entropy.
This behavior is illustrated in Fig.\,\ref{fig:bloch_mpemba}, where three initial states belonging to the excited-state hemisphere display a clear inversion of their relaxation times despite their initial ordering with respect to the stationary state. 
The results are obtained using the analytical solution of Eq.\,\eqref{Master equation}
(see End Matter for further details).

\begin{figure}[t]
    \includegraphics[width=\linewidth]{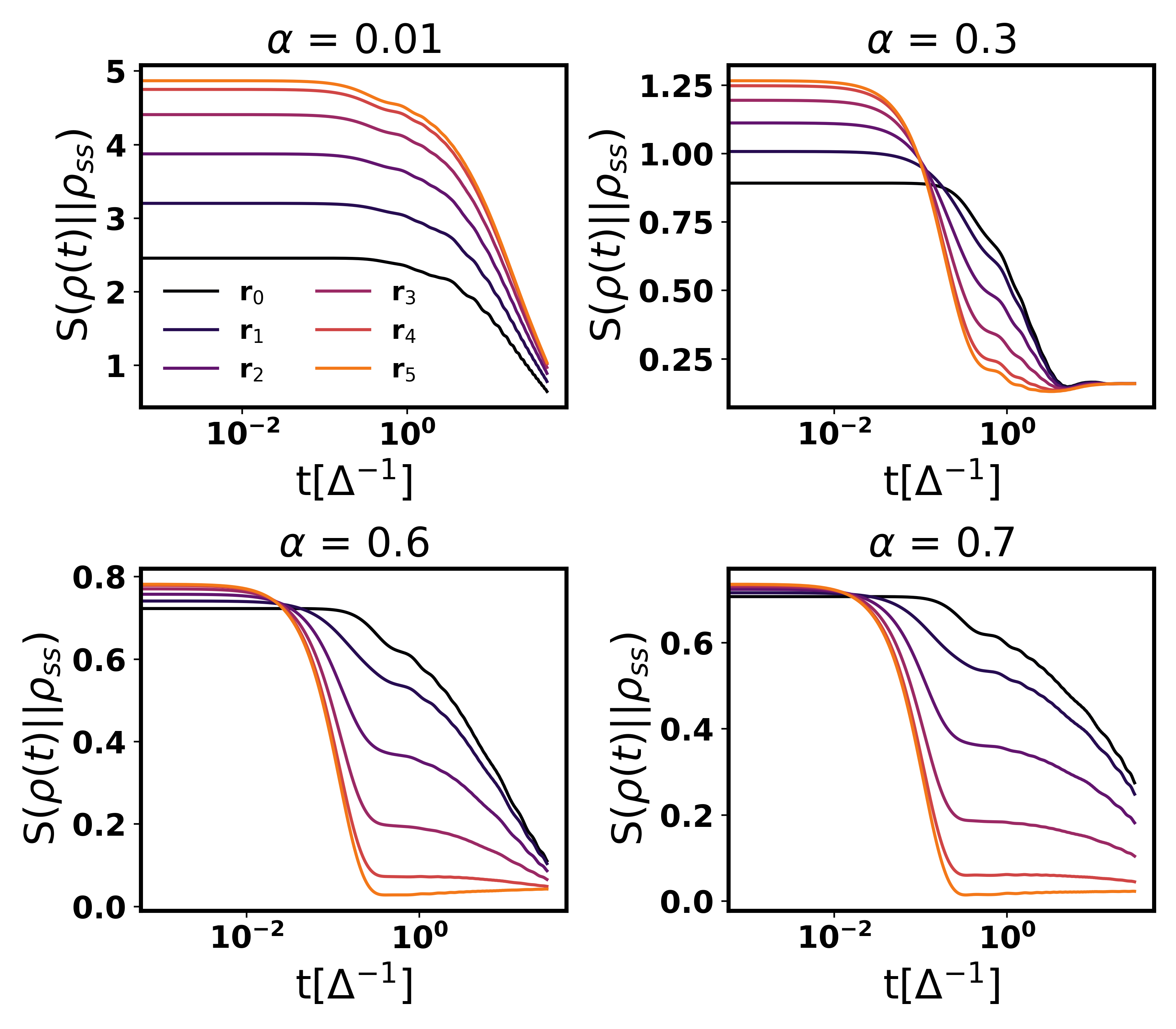}
        \caption{Quantum relative entropy with respect to the stationary state as a function of time for six initial states of the spin-boson model at zero temperature, labeled by the qubit Bloch vectors $\mathbf{r_i} = (-\mathrm{sin}(\phi_i),0,-\mathrm{cos}(\phi_i))$, with $\phi_i\in \{0,0.1\pi,0.2\pi, 0.3 \pi, 0.4\pi,0.5\pi\}$.
        (a) The curves of the quantum relative entropies do not intersect, signaling the absence of the effect in the weak-coupling case at zero temperature. (b), (c), (d) show the re-emergence of the effect in the relative entropy. For the simulations, the following numerical parameter values were used: $N=600, \Delta = 1, \omega_c =60 {\Delta}$.}
    \label{fig:panel_relative_entropy}
\end{figure}
We now explore the quantum Mpemba effect at zero temperature, both within and beyond the weak-coupling regime using as distances not only the quantum relative entropy but also the trace distance.
In the weak-coupling regime, we find that the quantum Mpemba effect disappears when relaxation is quantified by the quantum relative entropy at zero temperature, as shown in Fig.\,\ref{fig:panel_relative_entropy}(a).

In this limit, the quantum relative entropy decays monotonically and no crossing between different initial states is observed.
Remarkably, this behavior changes qualitatively as the system-bath coupling strength $\alpha$ is increased.
Moving beyond the weak-coupling regime, we observe the reemergence of the Mpemba effect in the quantum relative entropy exhibiting a clear crossing at a finite time.
This demonstrates that, while the effect is suppressed in the weak-coupling 
zero-temperature limit, it can be restored by increasing the coupling to the
environment, highlighting the crucial role of strong coupling to the environment.

\begin{figure}[t]
\includegraphics[width=\linewidth]{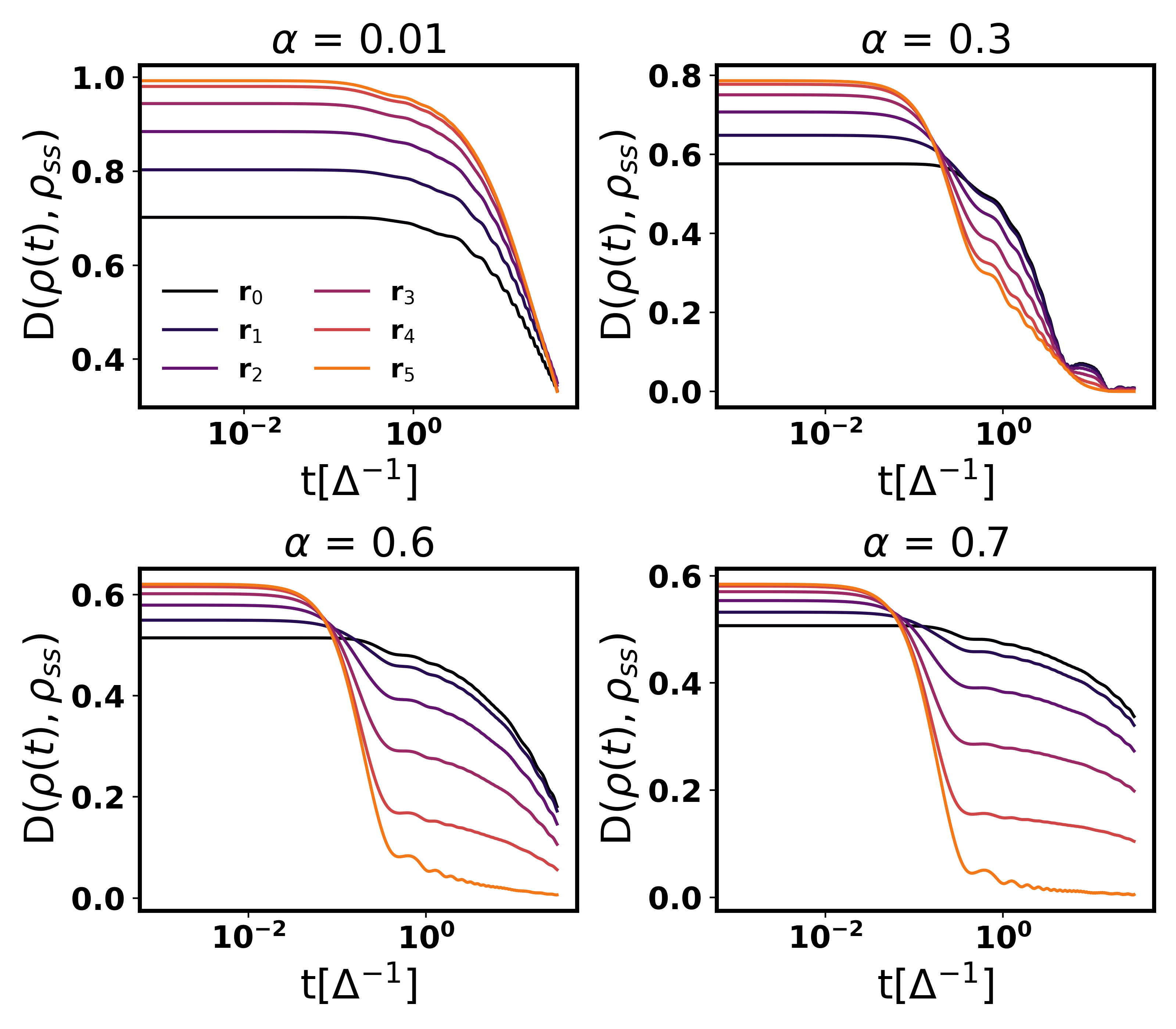}
\caption{Trace distance with respect to the stationary state as a function of time for different initial states at zero temperature, labeled by their Bloch vectors
$\mathbf{r_i} = (-\mathrm{sin}(\phi_i),0,-\mathrm{cos}(\phi_i))$, with $\phi_i\in \{0,0.1\pi,0.2\pi, 0.3 \pi, 0.4\pi, 0.5\pi\}$.
The effect occurs for any value of the coupling with the environment. For the simulations, the following numerical parameter values were used: $N=600,\Delta = 1, \omega_c =60 \Delta$.
}

\label{fig:panel_trace_distance}
\end{figure}
We performed the same analysis for the trace distance. 
As in the case of the quantum relative entropy, we observe an underlying geometric structure. In contrast to quantum relative entropy, the Mpemba effect can manifest when the trace distance is used as the distinguishability measure, even at zero temperature, within the weak-coupling regime, as shown in Fig.\,\ref{fig:panel_trace_distance}(a). With increasing $\alpha$, we observe that the effect persists and becomes more pronounced (faster relaxation for initial conditions farther from equilibrium as illustrated in Fig.\,\ref{fig:crossing_times_all} of the End Matter). 

We now discuss the physical mechanism underlying the enhancement of the Mpemba effect as $\alpha$ increases (see Figs.\,\ref{fig:panel_relative_entropy} and \ref{fig:panel_trace_distance}).

The enhancement of the Mpemba effect can be understood in terms of the separation between population and coherence relaxation timescales. For initial states in the excited-state hemisphere with the same Bloch-vector modulus, increasing the initial distance from the stationary state increases the population contribution to the distance while reducing the coherence contribution. Since the population component relaxes faster than the coherences, its contribution rapidly becomes negligible. The long-time dynamics is therefore dominated by the coherence components, which are smaller for the states that were initially farther from equilibrium. As a result, these states eventually become closer to the stationary state than the initially closer ones, producing an inversion of the relaxation ordering.
As the coupling strength approaches the critical value from below, $\alpha\rightarrow\alpha_c^{-}$	, the crossing time becomes increasingly short, as shown in Fig.\,\ref{fig:crossing_times_ohmic}. 
At the same time, the subsequent approach to the stationary state becomes slower. These two observations imply that near the transition, the dynamics develops a stronger separation of timescales, with a rapid initial decay of the population contribution followed by a slow long-time relaxation governed by the component associated with the critical mode. 
The resulting early-time inversion makes the Mpemba crossing effectively instantaneous, whereas critical slowing down controls the late-time relaxation.

To assess the robustness of this mechanism with respect to changes in the bath spectral density, we investigate how the relaxation dynamics and the quantum Mpemba effect depend on the spectral exponent $s$. We consider
two more cases: a sub-Ohmic bath with $s=0.2$ and a super-Ohmic bath with $s=1.5$.

\subsection{Sub-Ohmic regime}

We now consider a sub-Ohmic bath with spectral exponent $s = 0.2$, for which the delocalized-to-localized transition occurs at a critical coupling $\alpha_c \simeq 0.02$.
To facilitate comparison with the Ohmic case, we express the coupling in terms of the rescaled ratio $\alpha/\alpha_c$.

\begin{figure*}[!t]
    \centering

    \begin{subfigure}[t]{0.48\textwidth}
        \centering
        \includegraphics[width=\linewidth]
        {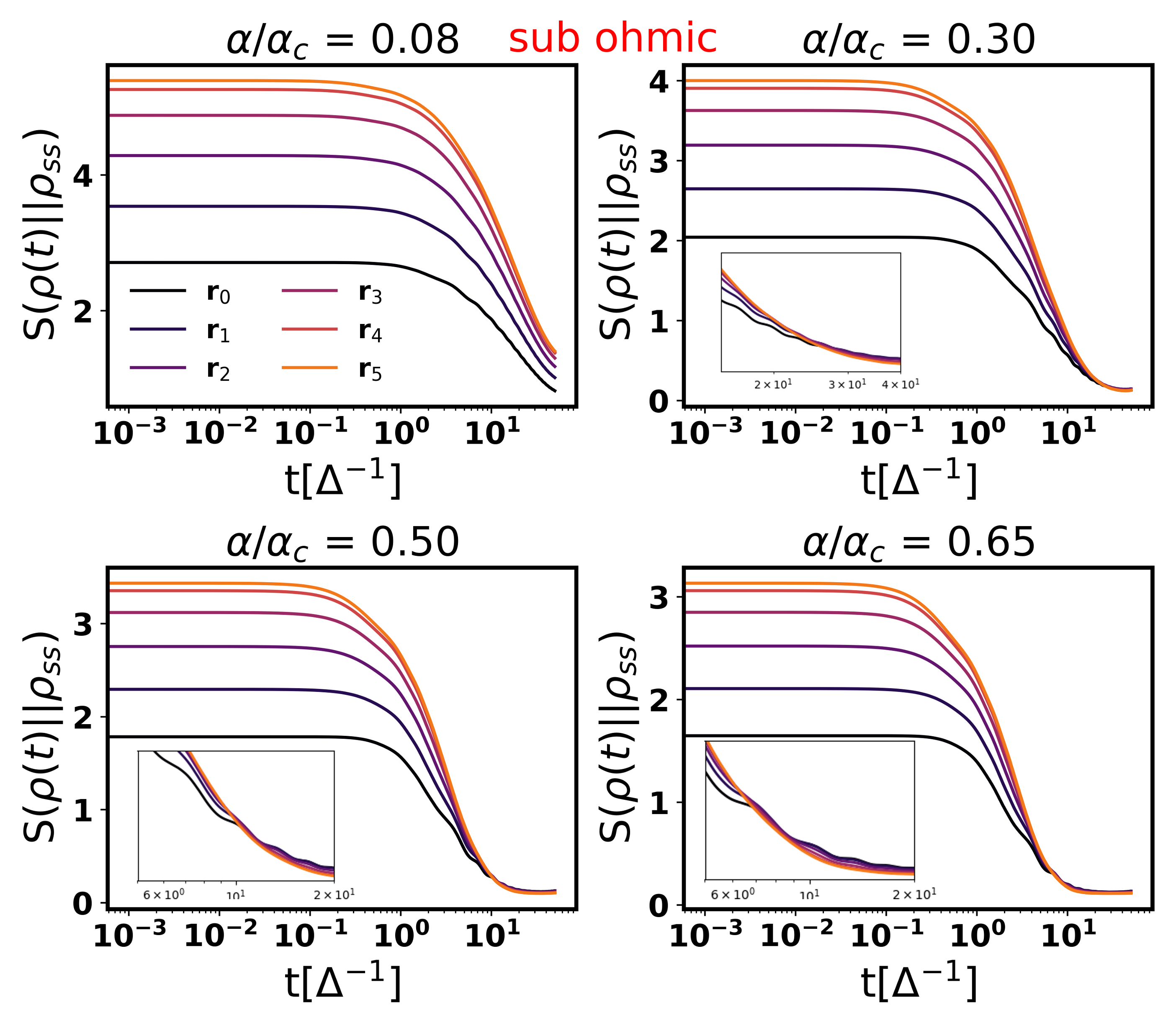}
        \caption{Sub-Ohmic regime: quantum relative entropy.}
        \label{fig:panel_relative_entropy_sub}
    \end{subfigure}
    \hfill
    \begin{subfigure}[t]{0.48\textwidth}
        \centering
        \includegraphics[width=\linewidth]
        {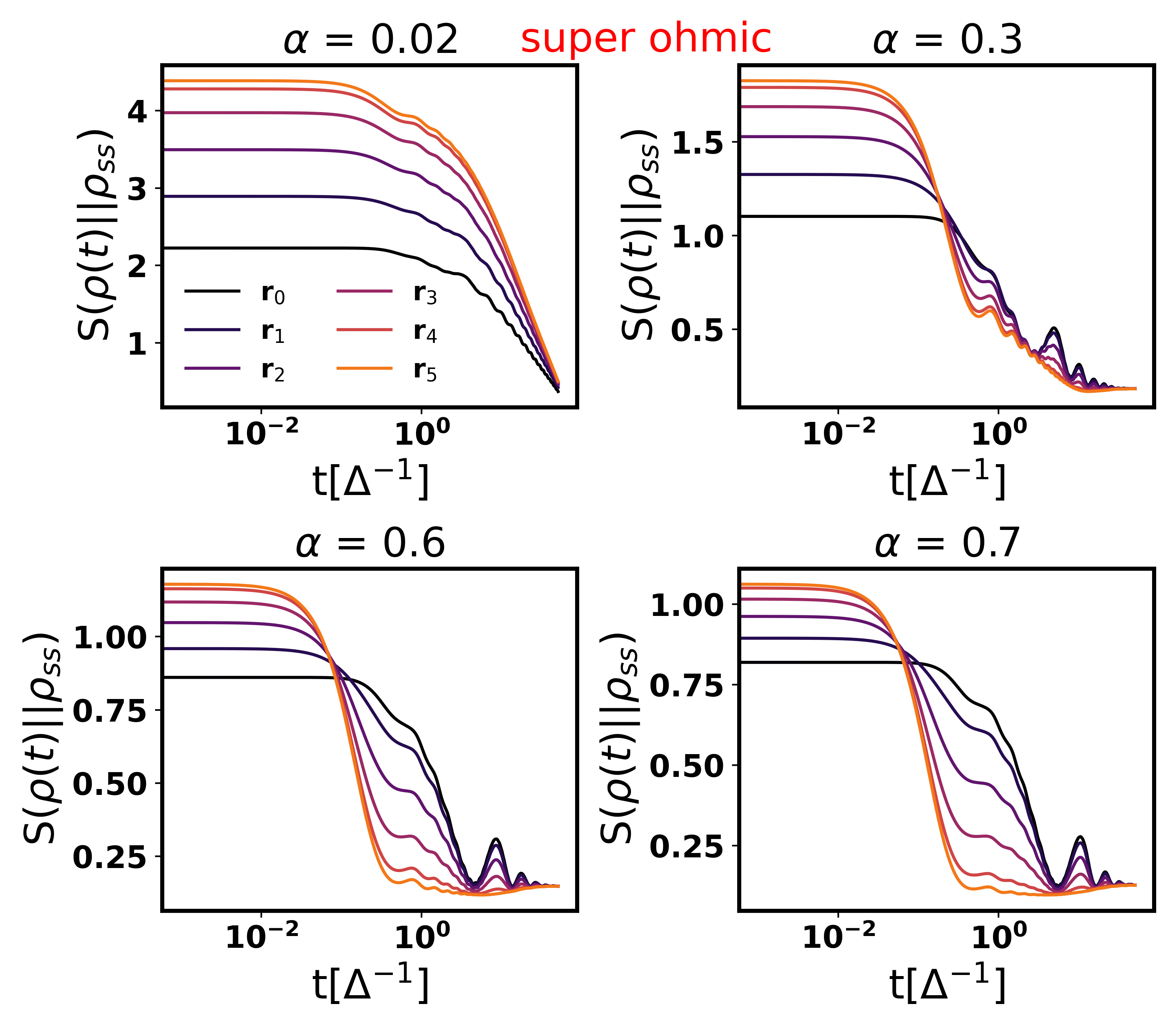}
        \caption{Super-Ohmic regime: quantum relative entropy.}
        \label{fig:panel_relative_entropy_super}
    \end{subfigure}

    \vspace{0.4cm}

    \begin{subfigure}[t]{0.48\textwidth}
        \centering
        \includegraphics[width=\linewidth]
        {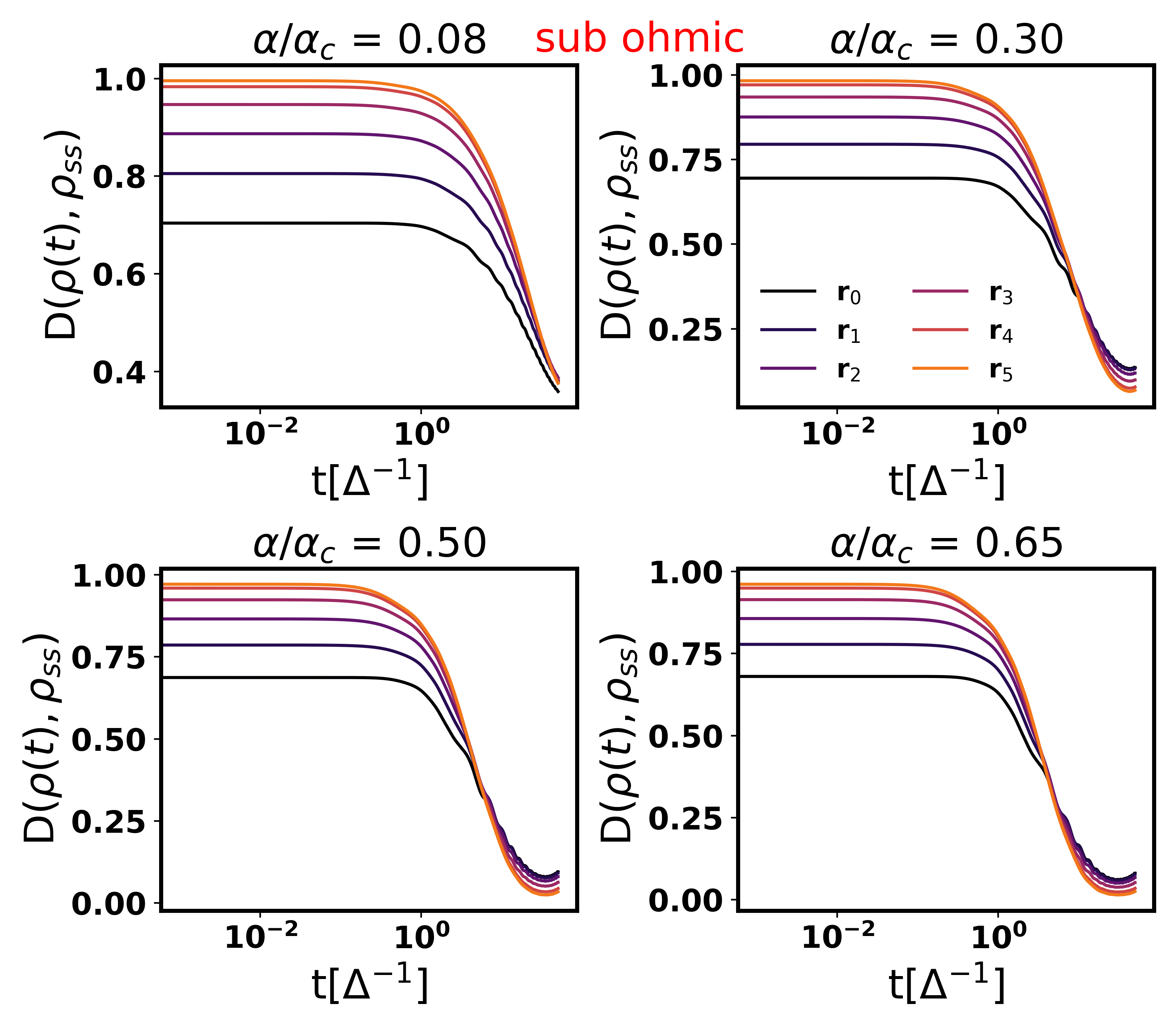}
        \caption{Sub-Ohmic regime: trace distance.}
        \label{fig:panel_trace_distance_sub}
    \end{subfigure}
    \hfill
    \begin{subfigure}[t]{0.48\textwidth}
        \centering
        \includegraphics[width=\linewidth]
        {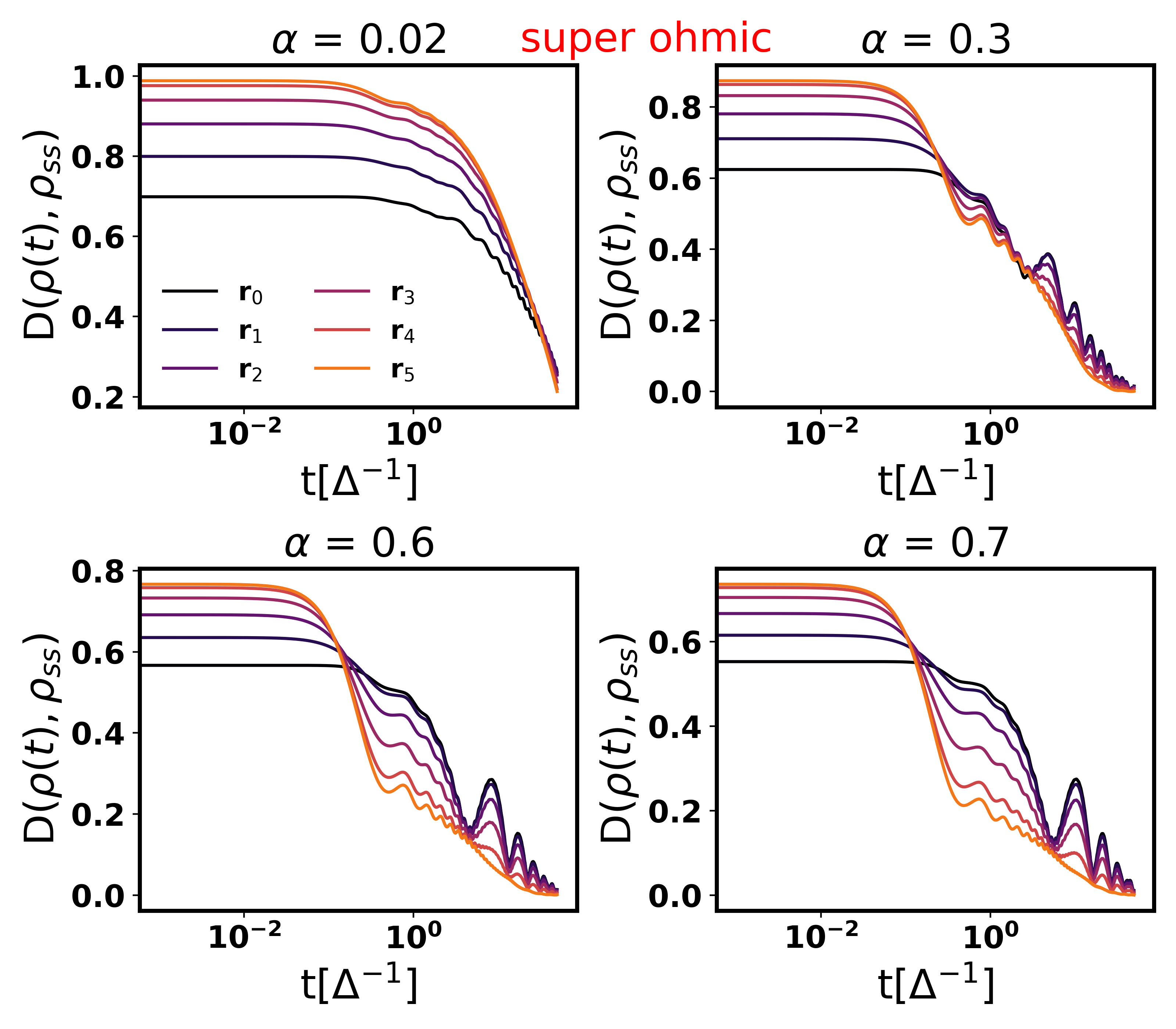}
        \caption{Super-Ohmic regime: trace distance.}
        \label{fig:panel_trace_distance_super}
    \end{subfigure}

    \caption{
    Relaxation dynamics in the sub-Ohmic and super-Ohmic regimes at zero
    temperature for six initial states of the spin--boson model,
    characterized by the Bloch vectors
    $\mathbf{r}_i=(-\sin\phi_i,0,-\cos\phi_i)$, with
    $\phi_i\in\{0,0.1\pi,0.2\pi,0.3\pi,0.4\pi,0.5\pi\}$.
    Panels (a) and (b) show the quantum relative entropy with respect to
    the stationary state in the sub-Ohmic and super-Ohmic regimes,
    respectively. At the weakest coupling, the curves preserve their
    initial ordering, whereas increasing the system--bath coupling leads
    to the re-emergence of the quantum Mpemba effect.
    Panels (c) and (d) show the corresponding trace-distance dynamics.
    In this case, crossings are observed for all the coupling strengths
    considered, and the inversion occurs at progressively shorter times
    as the coupling is increased.
    For all simulations, $N=600$, $\Delta=1$, and $\omega_c=60\Delta$.
    }
    \label{fig:sub_super_dynamics}

\end{figure*}

Figures \ref{fig:panel_relative_entropy_sub} and \ref{fig:panel_trace_distance_sub} show the relaxation dynamics quantified by the quantum relative entropy and the trace distance, respectively.
At weak coupling, $\alpha/\alpha_c = 0.08$, the relative-entropy curves preserve their initial ordering over the simulated time window, and no Mpemba crossing is observed. As the coupling is increased, the ordering is inverted at finite times, showing that the Mpemba effect is also restored in the sub-Ohmic regime.
The crossing is already visible at $\alpha/\alpha_c = 0.3$ and occurs at progressively shorter times for $\alpha/\alpha_c = 0.5$ and $0.65$.

The trace distance displays qualitatively similar but more robust behavior. A crossing is present for all the coupling strengths
considered, including the weakest-coupling case. 
Moreover, increasing $\alpha/\alpha_c$ shifts the inversion toward earlier times and makes the separation between the relaxation curves more
pronounced. 
These results show that the coupling-induced enhancement of the quantum Mpemba effect is not specific to the Ohmic bath and persists in the presence of a sub-Ohmic low-frequency spectrum.
\subsection{Super-Ohmic regime}
We now consider a super-Ohmic bath with spectral exponent $s = 1.5$, for which no phase transition occurs. We therefore characterize the dynamics directly in terms of the coupling strength $\alpha$.

Figures \ref{fig:panel_relative_entropy_super} and \ref{fig:panel_trace_distance_super} show the relaxation dynamics quantified by the quantum relative entropy and the trace distance, respectively.
At weak coupling, $\alpha = 0.02$, the relative-entropy curves preserve their initial ordering over the simulated time window, and no Mpemba crossing is observed. As the coupling is increased, the ordering is inverted at finite times, showing that the Mpemba effect is also restored in the super-Ohmic regime.
The crossing is already visible at $\alpha = 0.3$ and occurs at progressively shorter times for $\alpha = 0.6$ and $0.7$.

As above, the trace distance displays qualitatively similar but more robust behavior. A crossing is present for all the coupling strengths
considered, including the weakest-coupling case. 
Moreover, increasing $\alpha$ shifts the inversion toward earlier times and makes the separation between the relaxation curves more
pronounced. 
These results show that the coupling-induced enhancement of the quantum Mpemba effect is not specific to the Ohmic bath and persists in the presence of a super-Ohmic environment.
A distinctive feature of the super-Ohmic regime is the persistence
of oscillations at comparatively large coupling strengths. Similar
oscillatory behavior can also be observed in the $\alpha=0.3$ panel of
Fig.~\ref{fig:panel_trace_distance}. In the Ohmic spin--boson model,
however, the Toulouse point at $\alpha = 1/2$ marks the crossover
beyond which coherent oscillations are suppressed
\cite{le2010quantum}. No analogous finite-coupling Toulouse point
exists in the super-Ohmic case, and the oscillatory dynamics can
therefore persist at larger values of $\alpha$.

\section{Conclusions}
In this work, we have investigated the quantum Mpemba effect in the spin-boson model, with particular emphasis on the Bloch-sphere picture, the choice of distance measure, and the strength of the system-bath coupling.
By analyzing the relaxation dynamics on the Bloch sphere, we have shown that the effect exhibits a simple geometric structure for all spin-bath couplings: when initial states are restricted to the excited-state hemisphere, the Mpemba effect occurs for any pair of distinct initial states with the same Bloch-vector modulus.

We have characterized the relaxation process using both the trace distance and the quantum relative entropy.
While the Mpemba effect is clearly observed in the trace distance, we find that, at zero temperature and within the weak-coupling Markovian regime, the effect is absent when relaxation is quantified by the quantum relative entropy.
This highlights the sensitivity of the phenomenon to the choice of distance measure
and clarifies the thermodynamic nature of the effect in the Markovian limit.
By means of numerical simulations, we have shown that the Mpemba effect persists at strong system-bath coupling.
Remarkably, increasing the coupling strength leads to a reemergence of the effect also in the quantum relative entropy, indicating that strong-coupling effects play a crucial role in the relaxation dynamics.
Furthermore, by extending the analysis to sub-Ohmic and super-Ohmic
environments, we have shown that the observed behavior is robust with
respect to changes in the bath spectral properties. In particular, the
coupling-induced restoration of the Mpemba effect in the quantum
relative entropy and its persistence in the trace distance are not
restricted to the Ohmic case, but survive across qualitatively
different bath regimes.
Our results show that the population--coherence competition underlying the quantum Mpemba effect acquires a simple geometric structure in the spin--boson model and remains relevant beyond the weak-coupling Markovian regime. 
Increasing the system--bath coupling modifies this competition, enhancing the relaxation-order inversion and, for the quantum relative entropy, restoring a Mpemba crossing that is absent at zero temperature in the weak-coupling limit.
The persistence of this behavior across sub-Ohmic, Ohmic, and super-Ohmic environments further shows that the coupling-induced enhancement is not tied to the presence of a localization transition.
An important direction for future work is the extension of this analysis to dissipative many-body systems.
Spin chains coupled to quantum bosonic environments provide a natural setting for such an investigation, since the bath can strongly modify their equilibrium and dynamical properties and can even change the nature of the underlying quantum phase transition \cite{Perroni2023, 10.1093/acprof:oso/9780199213900.001.0001, le2010quantum}. 
It would therefore be interesting to investigate whether the geometric population--coherence mechanism identified here survives in interacting many-body systems and how collective modes, bath-induced correlations, and many-body criticality affect the occurrence and timescale of the quantum Mpemba effect.

\section*{Author contributions}

All authors contributed to the development of the project, the interpretation of the results, and the preparation of the manuscript.
P.C. and G.D.B. performed the numerical calculations and prepared the figures.
G.D.F. and C.A.P. supervised the project. All authors reviewed and approved the final
manuscript.

A large language model was used for language editing and proofreading
of the paper. All scientific content, calculations, results, and
conclusions were produced and verified by the authors.

\section*{Acknowledgements}
G.D.F. and C.A.P. acknowledge financial support from PNRR MUR Project No. PE0000023-NQSTI. G.D.B. and C.A.P. acknowledge funding from IQARO (Spin-orbitronic Quantum Bits in Reconfigurable 2D-Oxides) project of the European Union’s Horizon Europe research and innovation programme under grant agreement n. 101115190.

\appendix

\section{Analytical proof of the geometric structure of the quantum Mpemba effect}

In this End Matter we provide the analytical derivation supporting the results discussed in the main text concerning the geometric origin of the quantum Mpemba effect when relaxation is quantified by the trace distance. We focus on a two-level system in the weak-coupling Markovian regime, consistently with the assumptions adopted in the main text.

After a rotation to the energy eigenbasis, the system Hamiltonian in Eq.\,\eqref{Hamil_S} can be written as
\begin{equation}
H_S = \frac{\Omega}{2}\sigma_z .
\end{equation}
with $\Omega=\Delta$. 
The total Hamiltonian reads
\begin{equation}
H = H_S + H_B
+ \sigma_x \otimes \sum_k h_k^x (b_k + b_k^\dagger)
+ \sigma_z \otimes \sum_k h_k^z (b_k + b_k^\dagger).
\end{equation}
Although in the main text we focus on the spin-boson model with a single coupling operator (we consider the so-called amplitude damping), the analytical derivation presented here is carried out for a more general system-bath interaction, including both transverse and longitudinal couplings. 
Importantly, the resulting Lindblad dynamics of the reduced two-level system is fully characterized by the population and coherence relaxation times, $T_{\rm pop}$ and $T_{\rm coh}$, respectively, and the proof of the quantum Mpemba effect depends only on their relative magnitude. The results therefore apply \emph{a fortiori} to the specific coupling considered in the main text.

In the weak-coupling limit, the reduced dynamics is governed by a Lindblad master equation, which translates into the following equations for the Bloch components:
\begin{equation}
\begin{cases}
\dot r_x = -(\frac{\Gamma}{2}(2n_\beta+1) +\Gamma_{\rm dep})r_x - \Omega r_y, \\
\dot r_y = \Omega r_x -(\frac{\Gamma}{2}(2n_\beta+1) +\Gamma_{\rm dep}) r_y, \\
\dot r_z = -\left[\Gamma(2n_\beta+1)\right](r_z-r_z^{\rm eq}) .
\end{cases}
\label{eq:bloch_lindblad}
\end{equation}
Here $\Gamma$ and $\Gamma_{\rm dep}$ are the population relaxation and dephasing rates, respectively, and $r_z^{\rm eq}$ denotes the stationary value of the longitudinal component. For notational simplicity, we now specialize to the zero-temperature
case, for which $n_\beta=0$ and $r_z^{\rm eq}=-1$.
The solution of Eq.\,\eqref{eq:bloch_lindblad} reads
\begin{equation}
\begin{cases}
r_x(t) &= e^{-t/T_{\rm coh}}\!\left[r_x(0)\cos(\Omega t)-r_y(0)\sin(\Omega t)\right],\\
r_y(t) &= e^{-t/T_{\rm coh}}\!\left[r_x(0)\sin(\Omega t)+r_y(0)\cos(\Omega t)\right],\\
r_z(t) &= e^{-t/T_{\rm pop}}\!\left[r_z(0)-r_z^{\rm eq}\right]+r_z^{\rm eq}.
\end{cases}
\label{eq:lindblad_solution}
\end{equation}

with
\begin{equation}
T_{\rm pop}^{-1}=\Gamma, 
\qquad
T_{\rm coh}^{-1}=\frac{\Gamma}{2}+\Gamma_{\rm dep}.
\end{equation}

To establish the quantum Mpemba effect, consider two initial states $\rho_1$ and $\rho_2$ such that
\begin{equation}
D(\rho_1(0),\rho_{\rm ss})>D(\rho_2(0),\rho_{\rm ss}).
\end{equation}

Since the trace distance is non-negative, it is convenient to work with its square and define
\begin{equation}
\Delta(t)=D^2(\rho_1(t),\rho_{\rm ss})-D^2(\rho_2(t),\rho_{\rm ss}).
\end{equation}

Using Eqs.\eqref{eq:trace_distance} and \eqref{eq:lindblad_solution}, one finds
\begin{equation}\begin{split}
    \Delta(t)=\frac{1}{4}\Big[e^{-2t/T_{\rm pop}}\!\left((r_{1z}-r_z^{\rm eq})^2-(r_{2z}-r_z^{\rm eq})^2\right) +\\
+e^{-2t/T_{\rm coh}}\!\left(r_{1\perp}^2-r_{2\perp}^2\right)
\Big],\end{split}
\end{equation}
where $r_{i\perp}^2=r_{ix}^2+r_{iy}^2$.

We restrict to pairs of states with equal Bloch-vector modulus $|\mathbf r_1|=|\mathbf r_2|=|\mathbf r|$ and introduce spherical coordinates,
\begin{equation}
\mathbf r = |\mathbf r|(\sin\phi\cos\theta,\sin\phi\sin\theta,\cos\phi).
\end{equation}

The function $\Delta(t)$ can be rewritten as
\begin{equation}
\Delta(t)=\frac{|\mathbf r|^2}{4}e^{-2t/T_{\rm coh}}\,G(t),
\end{equation}
with
\begin{equation}
\begin{aligned}
G(t) ={}& \sin^2\phi_1-\sin^2\phi_2 \\
&+e^{-2t(1/T_{\rm pop}-1/T_{\rm coh})}
\left[
\cos^2\phi_1-\cos^2\phi_2\right.\\
&\left.-2\frac{r_z^{\rm eq}}{|\mathbf r|}
(\cos\phi_1-\cos\phi_2)
\right].
\end{aligned}
\end{equation}

The following argument applies whenever the population relaxation is faster than the coherence relaxation,
$T_{\rm pop}<T_{\rm coh}$. At zero temperature, this condition corresponds to
$\Gamma_{\rm dep}<\Gamma/2$. In particular, for the spin-boson coupling considered in the main text, for which
$\Gamma_{\rm dep}=0$, one has $T_{\rm pop}<T_{\rm coh}$. Under this condition, $G(t)$ is a monotonically decreasing function. The condition $\Delta(0)>0$ implies $\phi_2>\phi_1$. Taking the long-time limit,
\begin{equation}
\lim_{t\to\infty}G(t)=\sin^2\phi_1-\sin^2\phi_2 .
\end{equation}

For $\phi_1,\phi_2\in[0,\pi/2]$, corresponding to initial states in the excited-state hemisphere, the limit is negative, implying the existence of a finite time $t^*$ such that $\Delta (t^*)=0$ and $\Delta (t) <0$ for $t>t^*$.
The crossing time can be obtained explicitly as
\begin{equation}
t^* =\frac{T'}{2}\ln\!\left[
\frac{
\cos^2\phi_1-\cos^2\phi_2
-2\frac{r_z^{\rm eq}}{|\mathbf r|}(\cos\phi_1-\cos\phi_2)
}{
\sin^2\phi_2-\sin^2\phi_1
}
\right],
\end{equation}
with $(T')^{-1}=T_{\rm pop}^{-1}-T_{\rm coh}^{-1}$.

This completes the analytical proof of the geometric origin of the quantum Mpemba effect for the trace distance.

\section{Estimation of crossing times}

An important quantity for characterizing the quantum Mpemba effect is
the crossing time $t^*$, defined as the time at which the distance curves
associated with two initial states intersect. As shown by the analytical
derivation, the crossing time depends on the selected pair of initial
states,
\begin{equation}
    t^* = t^*(\mathbf{r}_i,\mathbf{r}_j).
\end{equation}
In the following, we fix $\mathbf{r}_5$ as the reference state and
extract $t^*$ numerically for the pairs
$(\mathbf{r}_i,\mathbf{r}_5)$.
\begin{figure*}[!t]
    \centering

    \begin{subfigure}[t]{0.48\textwidth}
        \centering
        \includegraphics[width=\linewidth]
        {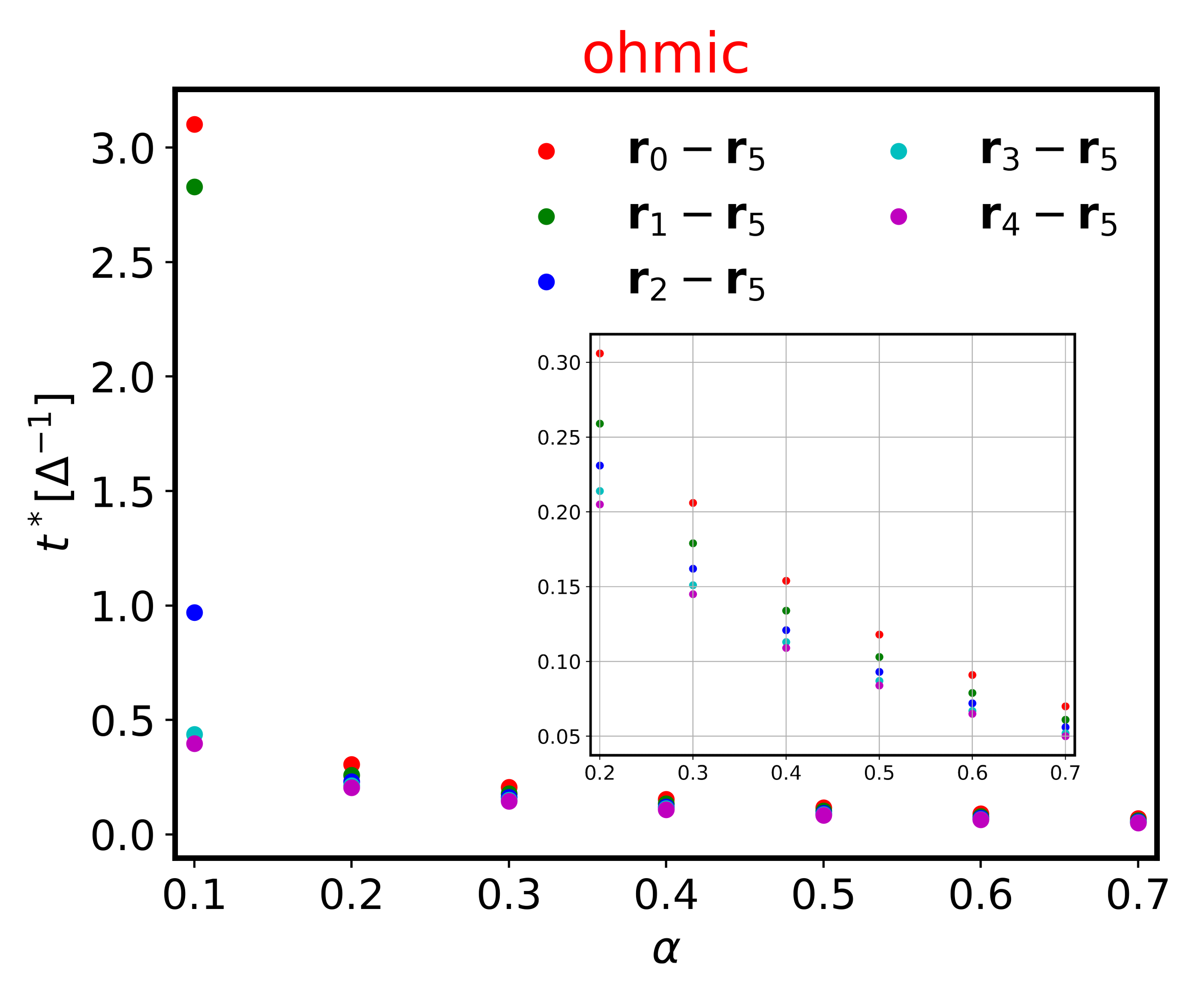}
        \caption{Ohmic regime.}
        \label{fig:crossing_times_ohmic}
    \end{subfigure}
    \hfill
    \begin{subfigure}[t]{0.48\textwidth}
        \centering
        \includegraphics[width=\linewidth]
        {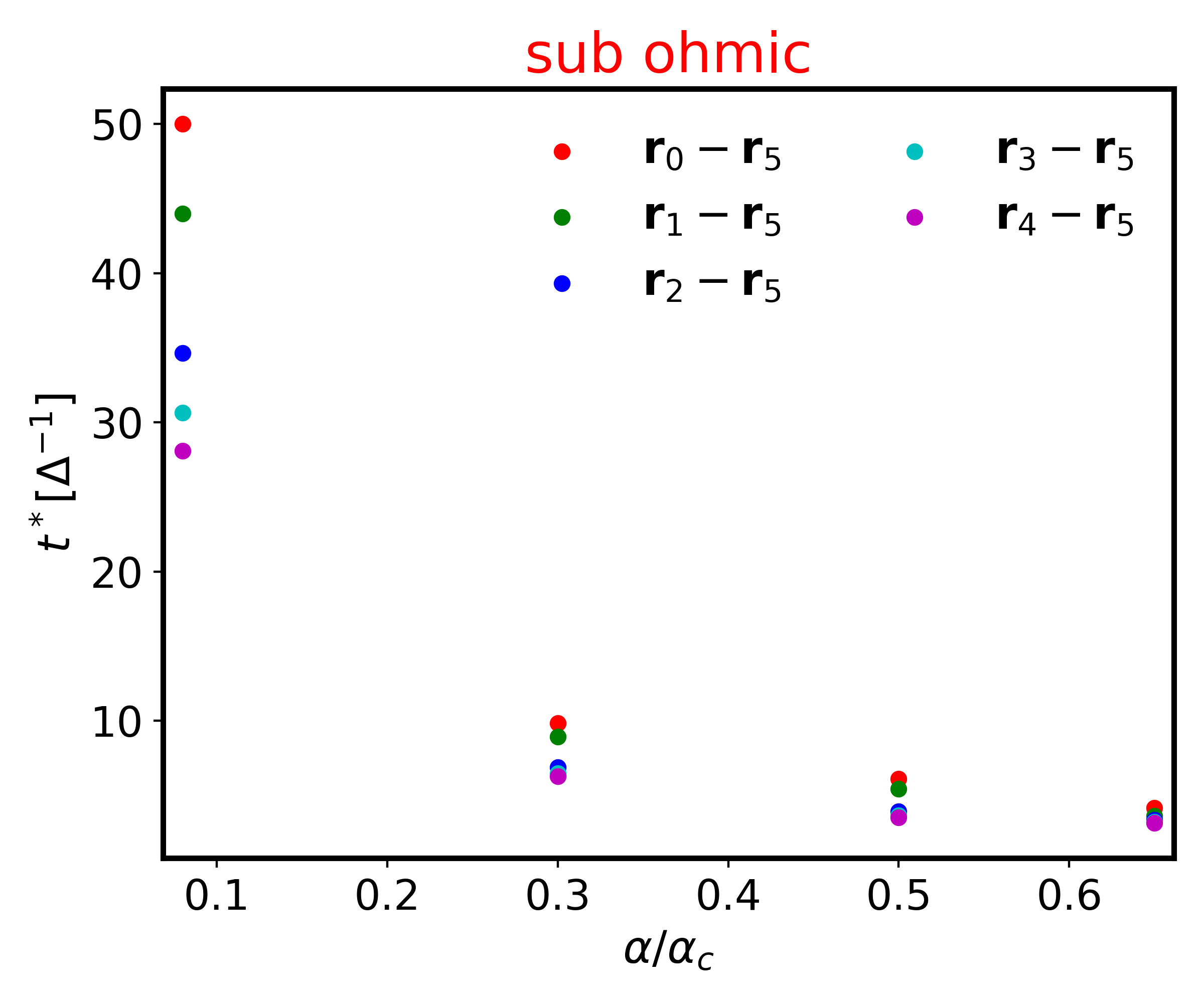}
        \caption{Sub-Ohmic regime.}
        \label{fig:crossing_times_sub}
    \end{subfigure}
    \par\vspace{0.3cm}
    \begin{subfigure}[t]{0.48\textwidth}
        \centering
        \includegraphics[width=\linewidth]
        {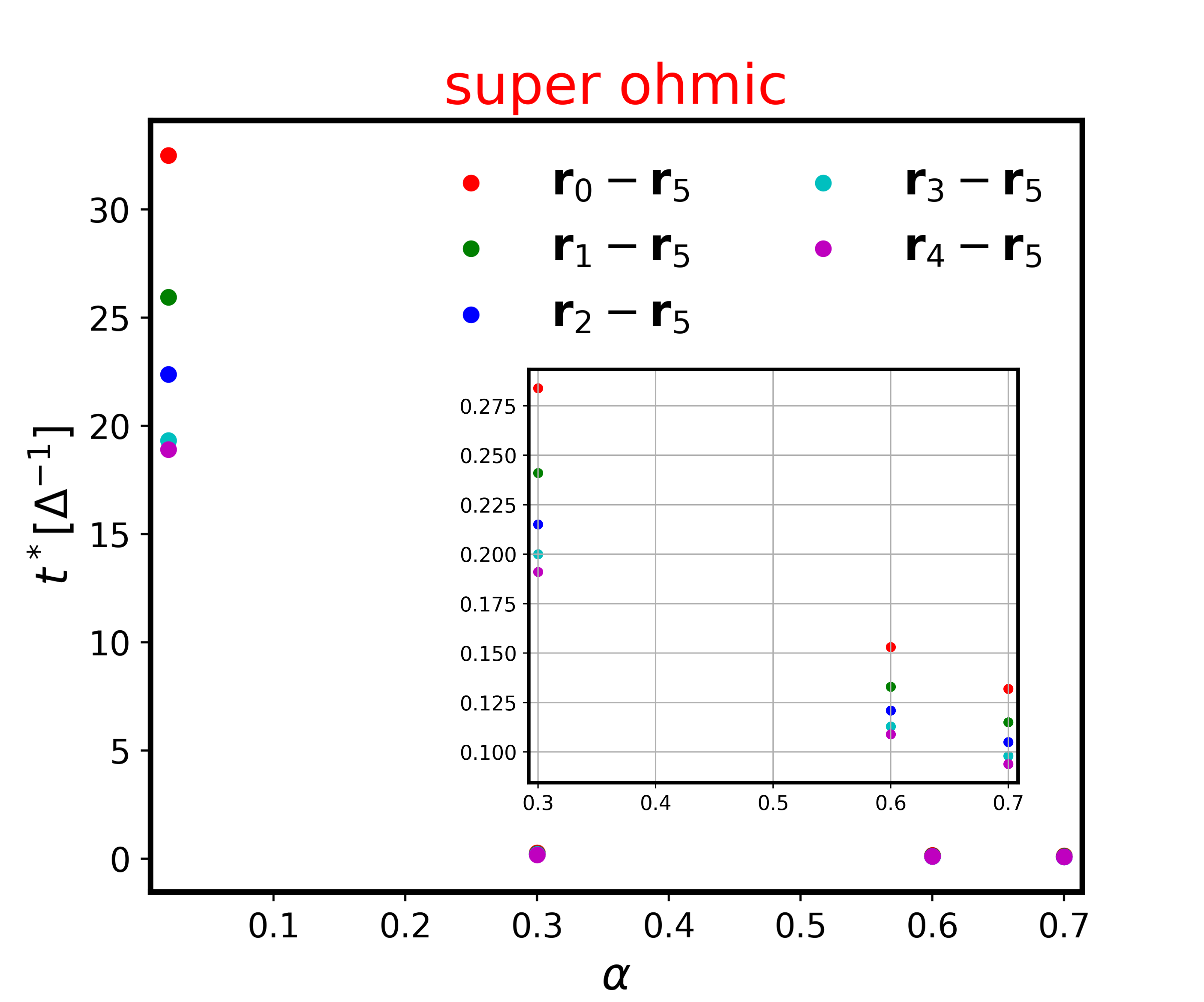}
        \caption{Super-Ohmic regime.}
        \label{fig:crossing_times_super}
    \end{subfigure}

    \caption{
    Crossing time $t^*$ of the trace distance for different pairs of
    initial states in the Ohmic, sub-Ohmic, and super-Ohmic regimes.
    The initial states are
    $\mathbf{r}_i=(-\sin\phi_i,0,-\cos\phi_i)$, with
    $\phi_i\in\{0,0.1\pi,0.2\pi,0.3\pi,0.4\pi,0.5\pi\}$, and
    $\mathbf{r}_5$ is taken as the reference state. Different symbols
    correspond to the pairs $(\mathbf{r}_i,\mathbf{r}_5)$, with
    $i=0,\ldots,4$.
    In the Ohmic and super-Ohmic
    cases, the coupling strength is reported as $\alpha$, whereas in the
    sub-Ohmic case it is expressed in terms of the rescaled coupling
    $\alpha/\alpha_c$.
    In all three regimes, the crossing time decreases as the
    system--bath coupling is increased, indicating that the inversion
    of the relaxation ordering takes place on progressively shorter
    timescales. The insets in panels (a) and (c) magnify the larger-coupling regions, where
    the crossing times become small.
    }
    \label{fig:crossing_times_all}
\end{figure*}

\begin{figure*}[!t]
    \centering

    \begin{subfigure}[t]{0.48\textwidth}
        \centering
        \includegraphics[width=\linewidth]
        {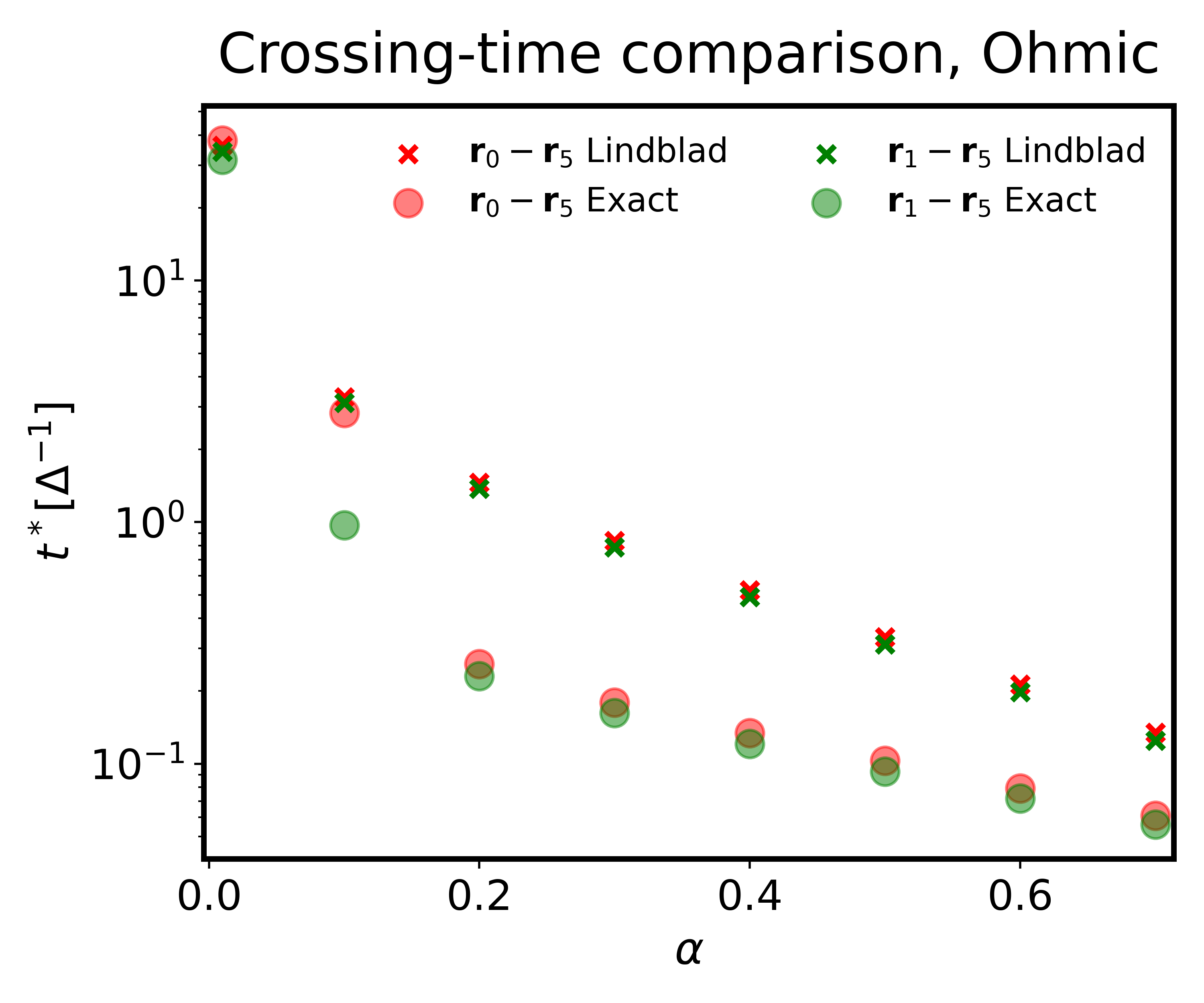}
        \caption{Ohmic regime.}
        \label{fig:crossing_comparison_ohmic}
    \end{subfigure}
    \hfill
    \begin{subfigure}[t]{0.48\textwidth}
        \centering
        \includegraphics[width=\linewidth]
        {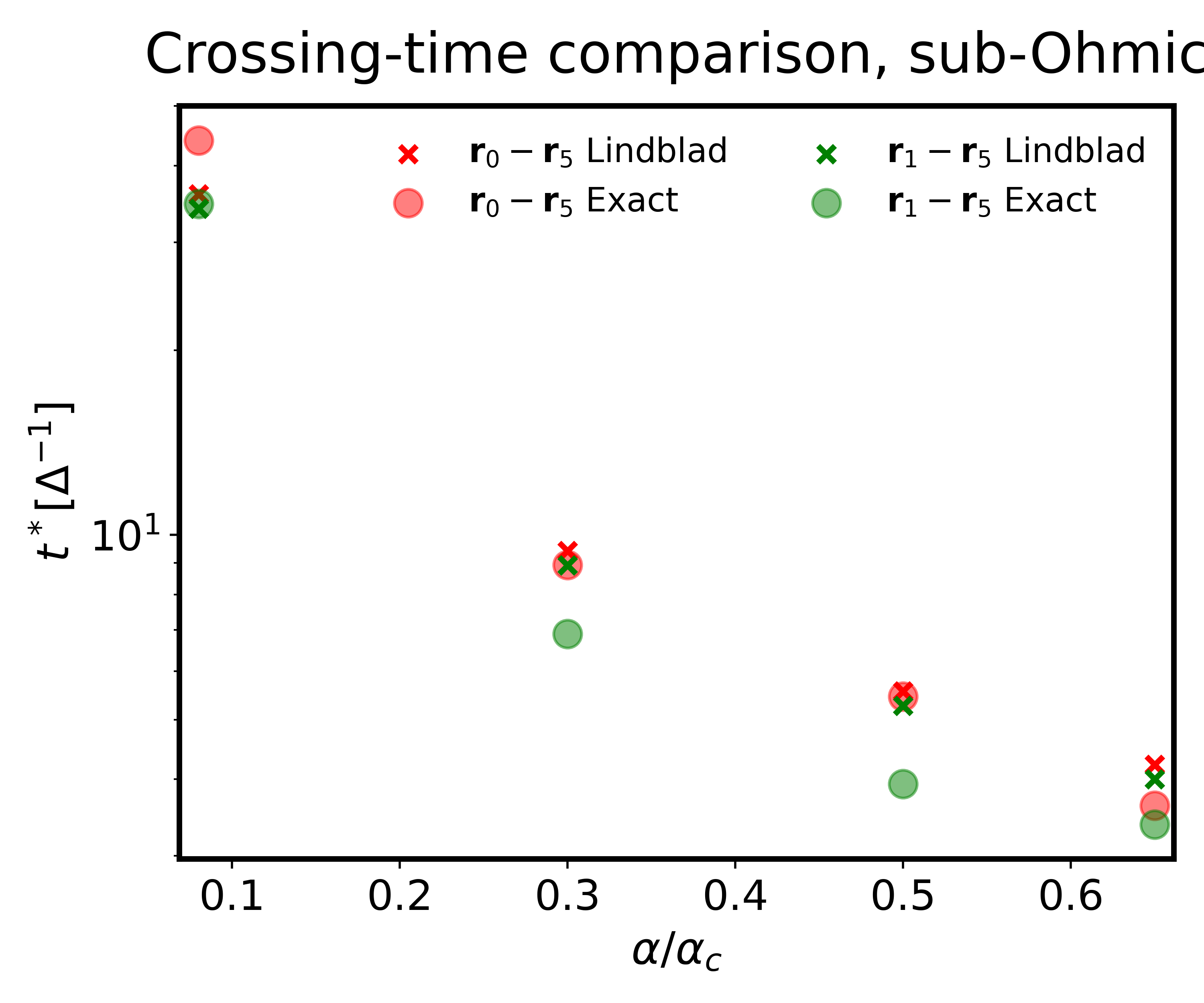}
        \caption{Sub-Ohmic regime.}
        \label{fig:crossing_comparison_sub}
    \end{subfigure}
    \par\vspace{0.3cm}
    \begin{subfigure}[t]{0.48\textwidth}
        \centering
        \includegraphics[width=\linewidth]
        {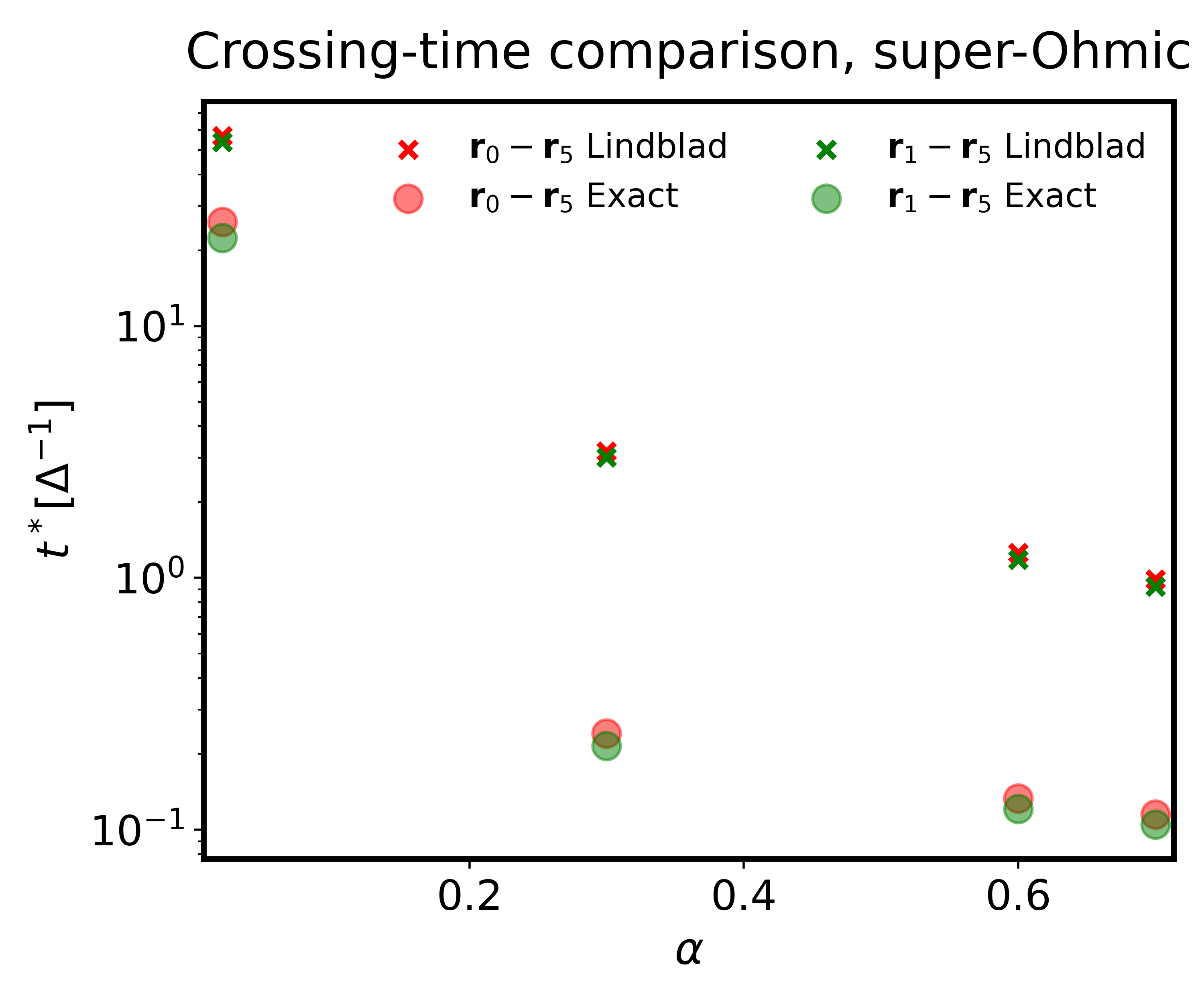}
        \caption{Super-Ohmic regime.}
        \label{fig:crossing_comparison_super}
    \end{subfigure}

    \caption{
    Comparison between the crossing times obtained from the exact
    system--bath dynamics (circles) and from the weak-coupling Lindblad
    prediction (crosses) for representative pairs of initial states.
    Panels (a), (b), and (c) correspond to the Ohmic, sub-Ohmic, and
    super-Ohmic regimes, respectively. In the Ohmic and super-Ohmic
    cases, the coupling strength is reported as $\alpha$, whereas in the
    sub-Ohmic case it is expressed in terms of the rescaled coupling
    $\alpha/\alpha_c$. The Lindblad prediction captures the qualitative
    decrease of the crossing time with increasing system--bath coupling,
    while quantitative deviations become increasingly visible beyond
    the weak-coupling regime. The discrepancy is particularly pronounced
    in the super-Ohmic case, where the Lindblad approximation
    substantially overestimates the exact crossing times at moderate and
    large coupling.
    }
    \label{fig:crossing_comparison}

\end{figure*}

The exact dynamics shows a systematic decrease of the crossing time as
the system--bath coupling is increased. This behavior is observed in
all the bath spectral regimes considered. In the Ohmic and sub-Ohmic
cases, $t^*$ becomes progressively shorter upon approaching the
critical region, whereas in the super-Ohmic regime a similar reduction
is found even in the absence of a finite-coupling localization
transition. Figure~\ref{fig:crossing_times_all} summarizes this behavior
for the different pairs of initial states considered. The suppression
of the crossing time is therefore not restricted to the vicinity of a
quantum critical point.
Furthermore, we directly compare the crossing times extracted from the
exact system--bath dynamics with the prediction obtained within the
weak-coupling Lindblad description. The comparison is shown in
Fig.~\ref{fig:crossing_comparison}, with panels (a), (b), and (c)
corresponding to the Ohmic, sub-Ohmic, and super-Ohmic regimes,
respectively for representative pairs of initial states.
In the Ohmic regime, the Lindblad prediction reproduces the
qualitative decrease of $t^*$ with increasing coupling, while
quantitative deviations become increasingly visible as the interaction
strength grows. The exact dynamics generally yields shorter crossing
times than the weak-coupling prediction.

A similar trend is observed in the sub-Ohmic regime. The Lindblad
description captures the overall reduction of $t^*$ with increasing
$\alpha/\alpha_c$, but the agreement is only quantitative over a
limited coupling range and depends on the selected pair of initial
states.

The largest deviations are found in the super-Ohmic regime. Although
both the exact and Lindblad dynamics predict a decrease of the crossing
time with increasing $\alpha$, the weak-coupling description
substantially overestimates $t^*$ once the interaction becomes
appreciable.

Overall, the comparison shows that the Lindblad theory captures the
qualitative coupling dependence of the crossing time, but becomes
progressively less accurate beyond the weak-coupling regime. The exact
dynamics exhibits an increasingly earlier inversion of the relaxation
ordering, highlighting the importance of finite-coupling effects in
setting the timescale of the quantum Mpemba effect.

\bibliographystyle{plainnat}
\bibliography{references}

\end{document}